\documentclass{aa}  

\usepackage{graphicx}
\usepackage{txfonts}
\usepackage{subcaption}
\usepackage{lscape}
\usepackage{placeins}

\usepackage[colorlinks=true, allcolors=blue]{hyperref}

\defcitealias{LimongiChieffi2018}{LC18}

\newcommand{\amnh}{Department of Astrophysics, American Museum of Natural History, 200 Central Park West, New York, NY 10024, USA}
\newcommand{\bath}{Department of Physics, University of Bath, Claverton Down, Bath, BA2 7AY, UK}
\newcommand{\surrey}{Department of Physics, University of Surrey, Guildford GU2 7XH, UK}
\newcommand{\columbia}{Department of Astronomy, Columbia University, New York, NY 10027, USA}
\newcommand{\uchicago}{Department of Astronomy \& Astrophysics, University of Chicago, 5640 S Ellis Ave, Chicago, IL 60637, USA}
\newcommand{\kicp}{Kavli Institute for Cosmological Physics, University of Chicago, Chicago, IL 60637, USA}
\newcommand{\lund}{Lund Observatory, Division of Astrophysics, Department of Physics, Lund University, Box 118, SE-221 00 Lund, Sweden}

\newcommand{\nsfsimons}{NSF-Simons AI Institute for the Sky (SkAI), 172 E. Chestnut St., Chicago, IL 60611, USA}
\newcommand{\cfa}{Center for Astrophysics | Harvard \& Smithsonian, Cambridge, MA 02138, USA}
\newcommand{\carnegie}{Carnegie Theoretical Astrophysics Center, Carnegie Observatories, 813 Santa Barbara St, Pasadena, CA 91101, USA}
\newcommand{\durham}{Institute for Computational Cosmology, Department of Physics, Durham University, South Road, Durham, DH1 3LE, UK}

\begin{document}

   \title{EDGE-INFERNO: The chemical signatures of assembly history in reionization fossils}
   \titlerunning{EDGE-INFERNO: The chemical signatures of assembly}
   
   \author{Eric~P.~Andersson\inst{1}
          \and Martin~P.~Rey\inst{2}
          \and Alexander~P.~Ji\inst{3,4,5}
          \and Justin~I.~Read\inst{6}
          \and Oscar~Agertz\inst{7}
          \and Alexander~H.~Riley\inst{7}
          \and Stacy~Y.~Kim\inst{8}
          \and Astha~Chaturvedi\inst{6}
          \and Jennifer~Mead\inst{9}
          \and Kaley~Brauer\inst{10}
          \and Mordecai-Mark Mac Low\inst{1}
          \and Andrew Pontzen\inst{11}
          }
          
    \institute{\amnh\\
              \email{eandersson@amnh.org}
              \and \bath
              \and \uchicago
              \and \kicp
              \and \nsfsimons
              \and \surrey
              \and \lund
              \and \carnegie
              \and \columbia
              \and \cfa
              \and \durham
             }

   \date{Received Month Day, Year}

  \abstract
  {Elemental abundances preserve a fossil record of the chemical environments in which stars form. In galaxies, stellar populations represent a superposition of both in situ and accreted components, tracing internal chemical evolution as well as the histories of merged satellites. In this work, we investigate how assembly history shapes the chemical properties of present-day stellar populations in low-mass dwarf galaxies. We analyze a suite of \textsc{edge-inferno} star-by-star cosmological zoom-in simulations of galaxies with present-day halo mass $\sim1.4\times10^9\,{\rm M}_{\odot}$. By applying the genetic modification technique to systematically vary galaxy growth while preserving the present-day halo mass, our simulations span a continuum of assembly histories, with the single-progenitor and multi-progenitor regimes representing its two limiting cases. Early-forming systems form most of their stars in a single progenitor, with centrally concentrated star formation and a continuous enrichment history. In contrast, our late-forming systems assemble by accretion of multiple progenitors with their own stellar populations, resulting in fragmented star formation across several halos with independent chemical evolution tracks. These regimes produce qualitatively different chemical signatures: single-progenitor systems exhibit connected, well-defined trends in abundance space, while multi-progenitor systems show multiple, distinct stellar populations with limited overlap. We find that metallicity distribution functions provide initial constraints on assembly history for samples of at least $50$-$100$ member stars, but additional chemical dimensions are often required to disentangle overlapping populations. Multi-element abundance analysis significantly improves the recovery of distinct components, although robust reconstruction still requires larger samples compared to metallicity distributions. Fully separating in situ and accreted populations using abundances alone remains challenging; this likely requires kinematic information. Accreted stars can contribute substantially even at high metallicities in multi-progenitor systems, while external enrichment can cause accreted material to follow similar chemical trends as in situ stars, particularly in the single-progenitor regime. These results highlight the importance of combining chemical and dynamical information when interpreting stellar populations in dwarf galaxies.
 }

   \maketitle
   
\section{Introduction}\label{sec:introduction}

The chemical abundances of stars are a key observable to constrain galaxy and star formation. These abundances are set by the chemical conditions of gas in their natal cloud to within 0.1 dex \citep{Dotter+2017}. Surveying stars over a range of ages and metallicities therefore allows us to archaeologically infer the star formation and chemical enrichment history of a given galaxy (see \citealt{Maiolino&Mannucci2019} for a review). Since chemical enrichment is tightly coupled to stellar nucleosynthesis, this information is invaluable to constrain stellar evolution and the life-cycle of stars (e.g. \citealt{Kobayashi+2020}). 

However, this inference is complicated by the non-linearity of star formation and chemical enrichment during the galaxy formation process. Stars are the primary site of nucleosynthesis, but different chemical elements are produced by different stellar evolution channels, each with its own timescale \citep{Kobayashi+2020}. Furthermore, galaxies are not closed systems, but rather accrete both pristine and pre-enriched gas that can dilute existing chemical abundances in the interstellar medium (ISM) and eject galactic outflows that carry chemical elements in and out of galaxies at differential rates \citep{Emerick+2018}. Galaxies also form hierarchically, accreting smaller galaxies \citep{Klypin+2011, Somerville+2015}. As a result, the present-day chemical composition of any given galaxy is a superposition of stellar populations: some are native (i.e., born {\it in situ}), shaped by the star formation, feedback and metal cycle of that galaxy; some are accreted (i.e., born {\it ex situ}), shaped by the histories of the accreted galaxies. Leveraging the chemistry of stars to constrain star and galaxy formation thus requires us to pinpoint the respective roles of each of these mechanisms. 

A particularly interesting regime in which to undertake such chemical inference is provided by faint dwarf galaxies (${\rm M}_{\star} \lesssim 10^6 \,  {\rm M}_{\odot}$, ${\rm L}_V \lesssim 10^5 \, {\rm L}_{\odot}$), for two reasons. First, these small dwarf galaxies formed most of their stars more than 10 billion years ago (\citealt{Brown+2014, Weisz+2014, Savino+2025}), before their star formation was permanently quenched by the ultraviolet background radiation irradiating the Universe after cosmic reionization (\citealt{Efstathiou1992, Shapiro+1994, Gnedin+2000, Bovill&Ricotti2009}). They thus have a short window for internal chemical evolution, forming only a few generations of stars, all of which are metal-poor ($\langle {\rm[Fe/H]} \rangle \lesssim -2$, where $\langle {\rm[Fe/H]} \rangle$ is the average iron metallicity of the ultra-faint dwarf). This, in theory, greatly simplifies the star formation history, increasing sensitivity to the underlying stellar evolution and associated nucleosynthesis. Second, their shallow gravitational potentials make them highly sensitive to the details of metal retention in the ISM \citep{Agertz+2020, Prgomet+2022, Baumschlager+2025, Rey+2025}, again increasing sensitivity to how stellar evolution channels inject energy into the ISM alongside their metals and drive galactic outflows.

This sensitivity to stellar evolution at low metallicities has motivated large efforts to obtain high-precision chemical data in faint dwarf galaxies (e.g. \citealt{Kirby+2009, Frebel+2010, Norris+2010, Vargas+2013}; see \citealt{Frebel&Norris2015, Simon2019} for reviews). In recent years, these efforts have expanded the state of the art in multiple directions: (i) more ultra-faint dwarf galaxies have been surveyed (e.g. \citealt{Chiti+2022, Heiger+2024, Heiger+2026, Kvasova+2024, Pace+2025}), reaching $\geq$ 10 hosts with chemical data available; (ii) more stars per ultra-faint dwarf galaxy have been surveyed (e.g. \citealt{Fu+2022, Fu+2023, Fu+2024, Chiti+2023, Bruce+2023, Luna+2025, Sandford+2026, Cerny+2026}), allowing some metal distribution functions (MDFs) to now be sampled with $\geq 100$ stars; (iii) more chemical elements per star have been measured (e.g. \citealt{Ji+2016, Ji+2016a, Ji+2018, Ji+2020, Mead+2024, Hong+2025}), unlocking  information about additional stellar evolution channels. With these new datasets, quantitative constraints on star formation and stellar evolution are starting to emerge, for example quantifying the chemical mixing in faint dwarfs' ISM (e.g. \citealt{Alexander+2023, Hansen+2024}), the burstiness of their star formation (e.g. \citealt{Ting&Ji2025}), or the contribution of specific progenitor channels such as Type Ia supernovae (SNeIa; \citealt{Alexander+2025}), r-process sites (\citealt{Ji+2016, Ji+2016b, Simon+2023}) or metal-free, Population III stars (Pop.III; \citealt{Rey+2025b}).

Underlying most of these analyses is the assumption that the chemical enrichment history of ultra-faint dwarf galaxies is dominated by their internal star formation (i.e. they only have an in situ population). This is in part due to the assumed simplified formation history of reionization-limited dwarfs and to the fact that sample sizes and numbers of chemical elements, which are essential to distinguish between ex situ and in situ populations in e.g. the Milky Way, remain low in these systems. However, low-mass halos generally experience fewer episodes of star formation and retain fewer metals due to relatively shallow gravitational potentials \citep{Mead+2025a}, therefore making signals from accretion more important. Nonetheless, testing the importance and contribution of an ex situ population and the validity of chemical inferences is challenging from a modeling perspective. Despite the fact that the star formation history is limited to $z\geq 4$, dry mergers and accretion can occur over 14 billion years and still shape the $z=0$ observables of these systems (\citealt{Rey+2019}). Accounting for this requires cosmological modeling that can self-consistently capture the merger history of these systems, while still resolving the internal chemical enrichment events of their building blocks.

Hydrodynamical cosmological zoom simulations now offer such a modeling solution. They have reached sufficient resolution ($1$--$10\,{\rm pc}$) to resolve the formation of ultra-faint dwarf galaxies, their ISM, and their chemical enrichment while capturing their mass growth over the age of the Universe (e.g. \citealt{Onorbe+2015, Jeon+2017, Wheeler+2019, Agertz+2020, Rey+2020, Applebaum+2021, Gutcke+2021, Tarumi+2021, Taylor+2025, Rey+2025}). Such tools now enable us to test how different modeling assumptions impact chemical observables of faint dwarfs such as their metallicities and abundances (e.g. \citealt{Jeon+2017, Gandhi+2022, Prgomet+2022, Ko+2025,  Rey+2025}). However, cosmological zoom simulations typically treat `stars' as integrated stellar populations, leading to stellar resolution $300$--$1000\,{\rm M}_{\odot}$ and $100$--$1000$ stellar particles in a $z=0$ dwarf. Consequently, stars originating in different progenitor galaxies are sometimes represented by only a handful of stellar particles, making it difficult to reconstruct the assembly history from the resolved stellar population.

Recent progress in star-by-star models can remedy to this limitation. Star-by-star models sample the IMF, explicitly capturing \textit{when} and \textit{where} stars inject energy and heavy elements \citep{Andersson+2020, Andersson+2023, Steinwandel+2023}. Such models have provided important contributions to our understanding of galactic winds \citep{Hu2019, Andersson+2020, Andersson+2023, Steinwandel+2023}, clustering of stars \citep{Hirai+2021, Calura+2022, Hislop+2022, Gutcke2024}, chemical enrichment \citep{Emerick+2019, Emerick+2020, Mead+2025a}, and star formation in the early Universe \citep{Brauer+2025a, Brauer+2025b, Brown+2026}. Today, these models have matured in performance and are available in cosmological simulations of ultra-faint dwarf galaxies, either individually sampling just the short-lived massive stars (e.g. \citealt{Gutcke+2022, Ko+2025}), or also sampling long-lived low-mass stars that are observed by chemical surveys at $z=0$ (\citealt{Andersson+2025}). In the latter case, this not only  provides exquisite stellar resolution ($\approx 100,000$ stars per dwarf), but also allows us to directly link individual-star predictions with individual-star observations and guide the interpretation of observed systems \citep{Andersson+2025}.

In this work, we quantify how different cosmological merger histories are imprinted on the stellar abundances of samples of individual low-mass stars at $z=0$ using the new suite of \textsc{edge-inferno} star-by-star cosmological zoom simulations (\citealt{Andersson+2025}). In a companion study, we tested how the star-by-star stellar abundances of a single dwarf galaxy respond to changes of inputs in chemical enrichment (e.g. yields, delay-time distribution; \citealt{Andersson+2026}). In this paper, we focus on the role of the merger history in setting chemical observables of faint dwarfs, using a fixed stellar evolution model. To facilitate this task, we analyze four `genetically-modified' cosmological simulations of faint dwarf galaxies. The genetic modification technique (\citealt{Roth+2016, Rey+2018, Pontzen+2021, Stopyra+2021}) allows us to introduce a targeted, controlled change in the mass growth history of a reference system, while keeping all other key cosmological variables conserved (e.g. total halo mass at $z=0$, large-scale environment, merger and accretion geometry). We use this approach to gradually increase and decrease the accreted fraction of a galaxy (see also \citealt{Rey+2019, Rey+2023}), allowing us to isolate the impact of the merger history on chemical abundances. 

We describe our methodology in Sect.~\ref{sec:method}, describing the numerical method in Sect.~\ref{subsec:numerical_method} and our initial conditions in Sect.~\ref{subsec:simulations}. In Sect.~\ref{sec:growth-history}, we describe the growth history of our simulated galaxies, providing results that give important context for subsequent sections. We present our investigation of chemical enrichment during galaxy assembly in Sect.~\ref{sec:enriched_environment} and show the resulting chemical signatures of present-day stellar populations in Sect.~\ref{sec:stellar_properties}. Finally, we discuss our results and conclude in Sect.~\ref{sec:conclusion}.   

\section{Methodology}\label{sec:method}

\begin{figure}
    \centering
    \includegraphics[width=0.975\linewidth]{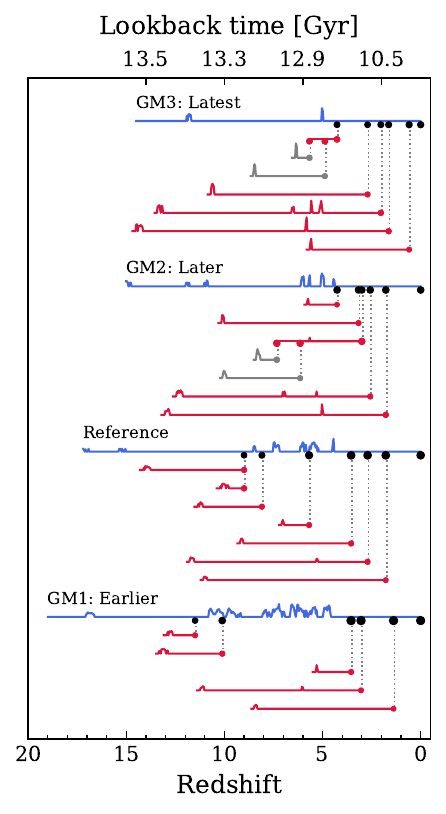}
    \caption{Merger trees of the four simulations investigated in this work, showing mergers at different redshift (or look-back time on the upper axis) with two points connected by a dotted gray line. Only halos containing stars are included. The size of the marker is scaled to the stellar mass of each galaxy prior to the merger. Using only in situ stars, the horizontal tracks show the star formation history of the main progenitor, main-progenitor satellites, and satellites of satellites, in blue, red, and gray respectively. The star formation rates have been rescaled by $\propto{\rm SFR}^{0.3}/\max(\rm SFR_{\rm sim})$ to emphasize small peaks (rates are quantified in Sect.~\ref{subsec:chemical_evolution}).}
    \label{fig:merger-history}
\end{figure}

\begin{table*}
    \centering
    \caption{Galaxy properties measured at redshift six and present day.}
    \begin{tabular}{l c c c c c c}
         &  $M_{200,\,z=6}$ & $M_{200,\,z=0}$ & $M_{\star,\,z=6}$ & $M_{\star,\,z=0}$, & $f_{{\rm in\, situ},\,z=6}$ & $f_{{\rm in\, situ},\,z=0}$\\
        Simulation &  [$10^9{\rm M}_{\odot}$] & [$10^9{\rm M}_{\odot}$] & [$10^5{\rm M}_{\odot}$] & [$10^5{\rm M}_{\odot}$] & [-] & [-] \\
        \hline\hline
        \texttt{GM1:Earlier} & 0.29 & 1.39 & 9.67 & 10.89 & 0.96 & 0.96 \\
        \texttt{Reference}   & 0.22 & 1.43 & 2.55 & 5.76 & 0.74 & 0.89 \\
        \texttt{GM2:Later}   & 0.09 & 1.43 & 0.48 & 1.67 & 0.25 & 0.60 \\
        \texttt{GM3:Latest}  & 0.03 & 1.37 & 0.03 & 0.93 & 0.06 & 0.15 \\
        \hline
    \end{tabular}
    \tablefoot{Halo mass, $M_{200}$, is defined by mass enclosed by the radius $R_{200}$ where the average density is 200 times the cosmic critical density. Stellar mass, $M_{\star}$, is measured within the same radius. The fraction of in situ stars, $f_{\rm in\ situ}$, is the relative fraction of stars that formed within $R_{200}$ of the main progenitor. From top to bottom, simulations are ordered by how quickly their dark matter halo assembles, controlled through engineered initial conditions (see main text for details).}
    \label{tab:h1459_properties}
\end{table*}

We applied the methodology of the \textsc{edge-inferno} framework, first presented in \citet{Andersson+2025}. Detailed descriptions of the numerical methods for the gas and $N$-body dynamics were given in \citet{Agertz+2020}; the initial conditions were described in \citet{Rey+2019}; star formation and feedback from individual stars were presented in \citet{Andersson+2023}; and the chemical enrichment model was detailed in \citet{Andersson+2026}. In the following section we summarize the most relevant aspects.

\subsection{Numerical methods and subgrid physics}\label{subsec:numerical_method}

The simulations were performed with the adaptive mesh refinement and $N$-body code \textsc{ramses}\footnote{https://ramses-organisation.readthedocs.io} \citep{Teyssier2002}. Gravity was computed by mapping particles onto the grid and solving Poisson’s equation using a multigrid method \citep{Guillet&Teyssier2011}. For the hydrodynamics solver, we assumed an ideal mono-atomic equation of state with an adiabatic index of 5/3. Between time steps, cooling and heating evolve the gas in the temperature range $1-10^9\,{\rm K}$ under equilibrium conditions for hydrogen and helium \citep{Courty&Alimi2004, Rosdahl+2013}, including metal-line cooling \citep{Ferland+1998}, and an evolving but spatially uniform UV background \citep{Haardt&Madau1996} suppressed by self-shielding \citep{Aubert&Teyssier2010, Rosdahl&Blaizot2012}. 

In the zoom-in region, the grid resolution was adapted to maintain a cell mass of approximately $167\,{\rm M}_{\odot}$, down to a maximum spatial resolution of $3.6\,{\rm pc}$. Dark matter particles in this region had a mass of $940\,{\rm M}_{\odot}$. Stars were treated as individual particles with masses determined by sampling the initial mass function (IMF) from \citet{Kroupa2001} over the range $0.08-100\,{\rm M}_{\odot}$. The IMF sampling followed the method of \citet{Sormani+2017}, implemented as in \citet{Andersson+2020}. To limit the total number of particles, stars with masses below $0.5\,{\rm M}_{\odot}$ were agglomerated into a single passive particle per star formation event. This mass threshold is low enough to ensure that all stars that contribute to feedback and enrichment during the age of the Universe are resolved particles. Furthermore, the agglomerated stars are not observable by modern spectroscopic facilities and are therefore ignored in most of our analyses \citep[for further details, see][]{Andersson+2025}.

Star formation was allowed in cells with densities exceeding $300\,{\rm cm}^{-3}$, following a Schmidt-like prescription in which the star formation rate density was proportional to the gas density divided by the local free-fall time, with an efficiency of $10\%$. At birth, stars inherited the velocity of their parent gas cell. In addition, their velocities were perturbed with a dispersion of $0.01\,{\rm km\,s}^{-1}$. Massive stars ($\geq8\,{\rm M}_{\odot}$) ejected fast ($1000\,{\rm km,s}^{-1}$) winds during the main sequence and ended as core-collapse supernovae (SNe) if their mass was below $25\,{\rm M}_{\odot}$, collapsing directly into black holes otherwise \citep{Janka2012, Zapartas+2021}. Low-mass stars remained passive during the main sequence, after which they ejected winds at a rate of $10^{-5}\,{\rm M}_{\odot}\,{\rm yr}^{-1}$ with a velocity of $10\,{\rm km\,s}^{-1}$ until their envelopes were ejected \citep{Hofner&Olofsson2018}. Main sequence lifetimes were calculated using the fitting functions of \citet{Raiteri+1996}, and all mass-loss rates were obtained via bilinear interpolation of the NuGrid yield tables \citep{Pignatari+2016, Ritter+2018}.

The feedback injection scheme followed the implementation of \citeauthor{Andersson+2023} (\citeyear{Andersson+2023}; see also \citealt{Agertz+2013}), treating stellar winds as sources of momentum and SNe as injections of thermal energy and momentum. For SNe, we assumed $10^{51}\,{\rm erg}$ of thermal energy and injected mass according to the same tables as for winds \citep{Pignatari+2016, Ritter+2018} in the case of core-collapse SNe. For SNeIa, explosion times were sampled from the time-delay distribution of \citet{Maoz&Graur2017}, with a minimum delay time of $38\,{\rm Myr}$ and normalized to a rate of $2.6\times10^{-13}\,{\rm yr}^{-1}\,{\rm M}_{\odot}^{-1}$. Mass loss and yields for SNeIa were interpolated from the tables of \citet{Seitenzahl+2013}. In cases where the cooling radius was unresolved ($<10\%$), the scheme switched to terminal momentum injection following \citet{Kim&Ostriker2015}. This approach has been shown to capture the total energy and momentum injection in our implementation \citep{Ohlin+2019}.

\subsection{Genetically engineered initial conditions}\label{subsec:simulations}

Using \textsc{genetIC} \citep{Roth+2016, Rey&Pontzen2018, Stopyra+2021}, our initial conditions were generated from the $\Lambda$ cold dark matter cosmology \citep[parameters from][]{Plank+2014}. A reference halo was selected with a present-day halo mass of $M_{200} = 1.4\times10^{9}\,{\rm M}_{\odot}$, defined as the mass enclosed with the radius where the mean density was 200 times the cosmic critical density. Three alternative growth histories for this halo was obtained by modifying the halo mass at redshift $z=6$, while maintaining the same mean density in the Lagrangian region (defined by particles that are inside twice the virial radius at $z=0$). In this way, the four galaxies have the same $z=0$ halo mass, but they grow at different rates, while keeping changes to the surrounding large-scale environment minimal \citep[for further details, see][]{Rey+2019}. We label our simulations using the convention of \citet{Andersson+2025}, that is, by whether they grow in halo mass earlier or later compared the simulations without modification, labeled as \texttt{Reference}. The labels for genetically modified simulations include the prefix \texttt{GM}.

\section{Growth histories and their impact on present-day observables}\label{sec:growth-history}

\begin{figure*}
    \centering
    \includegraphics[width=0.975\linewidth]{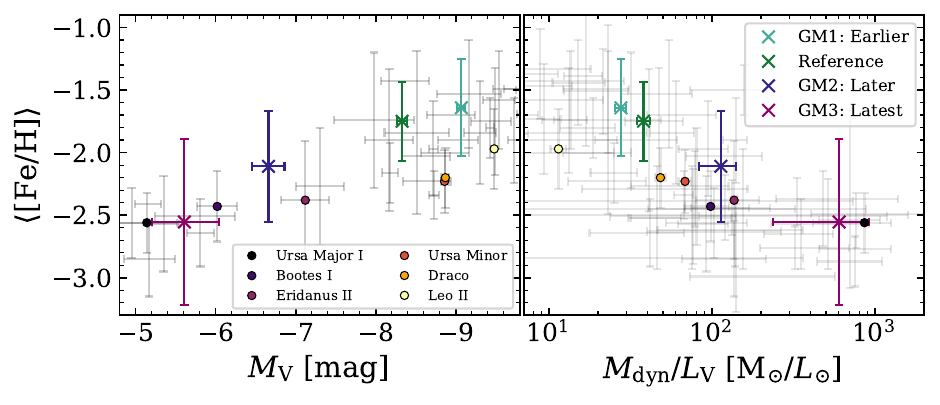}
    \caption{Mean of the [Fe/H] distribution as a function of V-band magnitude (left) and mass-to-light ratio (right) for our simulations shown in colors denoted in the legend. Grey error bars show galaxies in the Local Group collected from the database of \citet{Pace2025}, with a few examples highlighted for the discussion in Sect.~\ref{subsec:observations}. Vertical errors indicate the 68th percentiles of the [Fe/H] distribution (metal spread), while the horizontal errors indicate measurement uncertainty for observed galaxies and shot noise for simulated galaxies \citep[see][for details]{Andersson+2025}. The mass-to-light ratio emphasizes the disparity between halo mass and stellar mass caused by differences in assembly history, as is shown by the trend in our simulations where assembly history is engineered to give galaxies that form earlier and later compared to a reference.  As in \citet{Andersson+2025}, stars with [Fe/H]$<-4$ was omitted in our calculations to avoid possibly spurious effects due to missing Pop. III physics (see Sect.~\ref{sec:stellar_properties} for further details).}
    \label{fig:feh_MdynLV}
\end{figure*}

To compare our simulations and isolate the dependence on different growth history, we define the merger tree based on the \texttt{Reference} simulation. Using this simulation, we choose the main progenitor of the merger history by following the most massive halo in each merger back in time. In the other simulations, we instead define the merger history by visually inspecting the large-scale structure and ensuring that we follow the position of the halo that was genetically modified, i.e., the main progenitor in \texttt{Reference}. Note that this does not guarantee that the main progenitor in the engineered simulations follows the most massive halo in each merger.\footnote{For \texttt{GM1: Earlier} and \texttt{GM2: Later}, the merger tree follows the most massive halo in each merger event. At very early times, mergers in \texttt{GM3: Latest} have been modified to an extent where this is no longer the case. Instead, the galaxy forms out of halos with roughly equal mass, and a main progenitor is not robustly defined. After redshift 8, \texttt{GM3: Latest} has a merger tree that follows the most massive halo in each merger.}

We present the assembly histories as merger trees in Fig.~\ref{fig:merger-history} and quantify the differences in Table~\ref{tab:h1459_properties} by presenting the mass of the galaxies at $z=6$ and $z=0$. Figure~\ref{fig:merger-history} shows the redshifts when mergers occur in the different simulations, as well as the star formation rate of each system. Note that our initial condition modifications will introduce changes that broadly shifts mergers to occur either earlier or later relative to the reference simulation, as is indicated by how each simulation is labeled. Furthermore, most star formation is concentrated in a single halo in \texttt{GM1: Earlier}, and spread among and increasing number of systems in simulations placed further up in the diagram. We will come back to this in Sect.~\ref{sec:enriched_environment}. We refer to \citet{Andersson+2025} for further details about the halo growth rate and cumulative stellar mass of our simulations.

An important feature of these simulations is the systematic increase in stellar mass (at fixed present-day halo mass) for galaxies that assemble earlier \citep[see also][]{Rey+2019}. This trend arises from the near-complete quenching of star formation in low-mass galaxies ($\lesssim10^9\,\mathrm{M_\odot}$) during reionization \citep[][]{Bovill&Ricotti2009, Brown+2014, Weisz+2014}. Consequently, halo growth prior to reionization is a strong predictor of the final stellar mass.

A larger stellar mass also naturally leads to higher mean iron abundances due to the greater number of stars contributing to chemical enrichment. This gives rise to the well-known tight relation between stellar mass and metallicity \citep{Simon&Geha2007, Kirby+2013, Read+2017, Rey+2025}. We show this relation in terms of mean [Fe/H] as a function of total $V$-band magnitude in the left panel of Fig.~\ref{fig:feh_MdynLV}. Shown using gray error bars are estimates from observed systems using data from \citet{Carlin+2009, Charles+2023, Collins+2021, Heiger+2024, Jenkins+2021, Ji+2021, Kirby+2013, Kirby+2015, Kirby+2020, Kvasova+2024, Li+2017, Martin+2014, Muller+2025, Pace+2020, Simon2019, Slater+2015, Wojno+2020}, collected by \citet{Pace2025}. Even with galaxies at fixed present-day halo mass, our simulations follow the relation observed for Local Group dwarf galaxies. Notably, modifying the growth history moves galaxies significantly farther along this relation than internal effects such as stochastic sampling of the IMF, the timing of SNeIa, or different assumptions for nucleosynthetic yields \citep{Andersson+2026}.

Furthermore, because we fix the present-day halo mass, the systems show systematically different mass-to-light ratios. With an earlier and increased iron production at higher stellar mass, the impact of accretion history is emphasized in trends between mean [Fe/H] and mass-to-light ratio, shown in the right panel of Fig.~\ref{fig:feh_MdynLV}. For this figure, $M_{\rm dyn}$ is defined following \citet{Wolf+2010},
\begin{equation}\label{eq:Mdyn}
    M_{\rm dyn} = \frac{3\sigma^2r_{1/2}}{G},
\end{equation}
where $\sigma=\sigma_{\rm 3D}/\sqrt{3}$ is equivalent to the line-of-sight stellar velocity dispersion, $r_{1/2}$ is the half-light radius, and $G$ is the gravitational constant. Note that these quantities suffer from shot noise introduced by sampling the giant branch (see \citealt{Andersson+2025} for details). This noise depends on distance, density of background sources and instrument details. We include estimates of this noise as errors on our measurements in Fig.~\ref{fig:feh_MdynLV}, using the errors derived in \citet{Andersson+2025} for $M_{\rm V}$ and $r_{1/2}$. The same noise does not trivially apply to the $\sigma$, which typically requires spectroscopic follow-up on stars that trace the gravitational potential. The reported observations in Fig.~\ref{fig:feh_MdynLV} include measurement uncertainties on the velocity dispersion, typically around $0.5\,{\rm km\,s}^{-1}$ but reaching $3\,{\rm km\,s}^{-1}$ for more distant galaxies \citep{Pace2025}. We omit an equivalent uncertainty when calculating $M_{\rm dyn}$ for our simulations. To assess its potential impact, we repeated the error propagation assuming uncertainties of $0.5$ and $3\,{\rm km\,s}^{-1}$. The former increases the resulting $M_{\rm dyn}/L_V$ uncertainties by factors of $1.03$–$1.91$, while the latter increases them by factors of $1.85$–$9.83$. The effect is largest for systems with low velocity dispersions because $\sigma$ enters Eq.~\eqref{eq:Mdyn} quadratically.

Finally, we emphasize that these systems evolve dynamically without active star formation after $z=4$, allowing only ex situ stellar mass growth in the latter $\sim 12\,{\rm Gyr}$ of their evolution. The implication is that the chemical abundances found in stars at $z=0$ can be heavily influenced by accreted material, and as we will show next, the chemical enrichment of the larger environment around the galaxy progenitor.

\section{Chemical enrichment around galaxies with different assembly history}\label{sec:enriched_environment}

\begin{figure*}
    \centering
    \includegraphics[width=\linewidth]{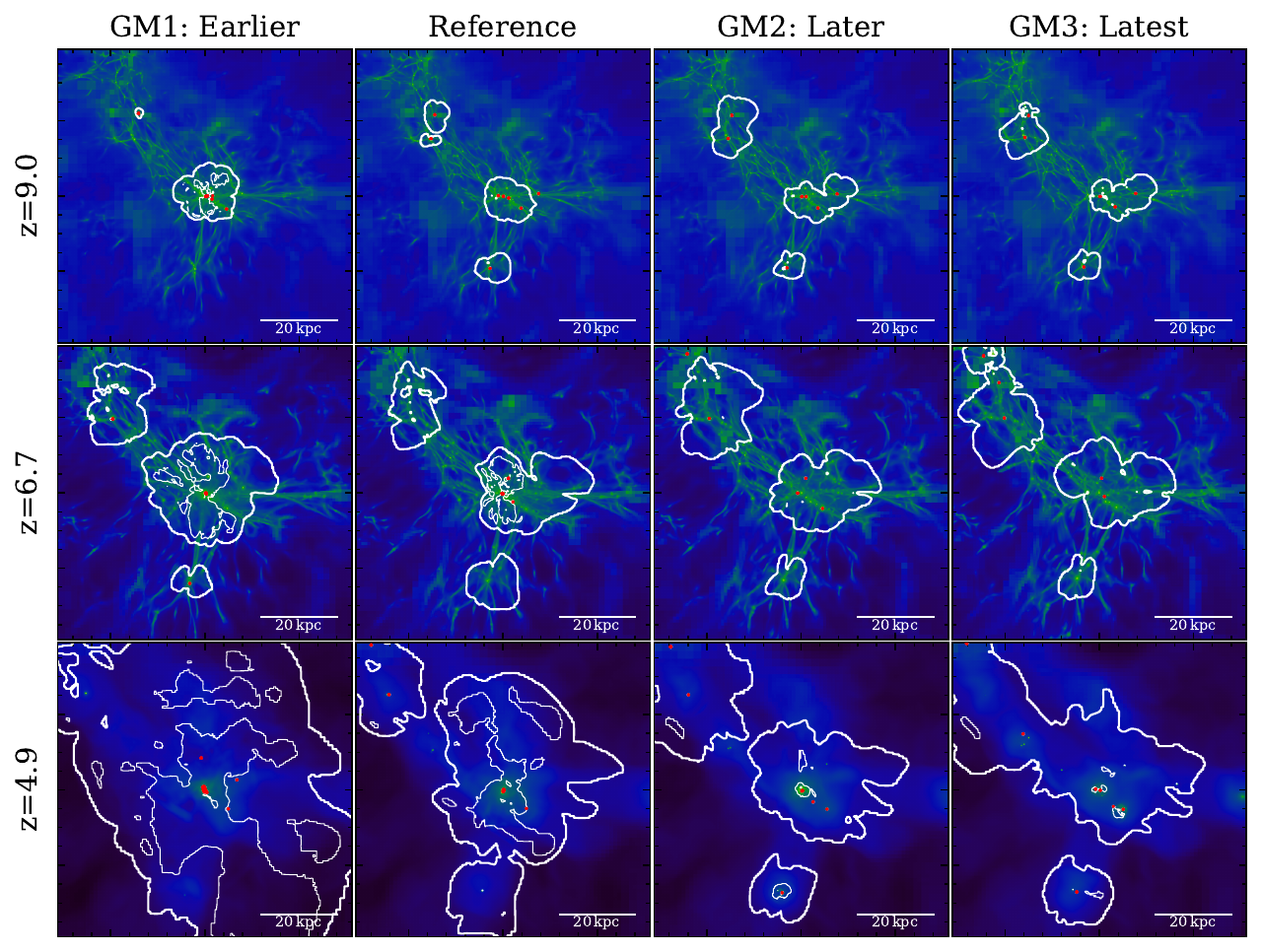}
    \caption{Projected gas density structure at redshifts 9.0, 6.7, and 4.9 (top, middle, and bottom row, respectively) for simulations with different growth histories (columns as labeled). Thick (thin) white contours show projected gas [Fe/H] at -4 (-2). Red points indicate the positions of newly formed stars, i.e., stars that spawned after the redshift of the previous row (the first row shows all stars formed at this point). In \texttt{GM1: Earlier}, a single progenitor is the predominant source of enriched material, while the galaxies where mass is assembled later show increasingly patchy regions of high enrichment. As a result, the volume filled by gas with $\rm [Fe/H]>-2$ grows significantly faster in simulations where star formation is concentrated in a single progenitor (e.g., \texttt{GM1: Earlier}).}
    \label{fig:halo_projection_map}
\end{figure*}

\begin{figure}
    \centering
    \includegraphics[width=0.95\linewidth]{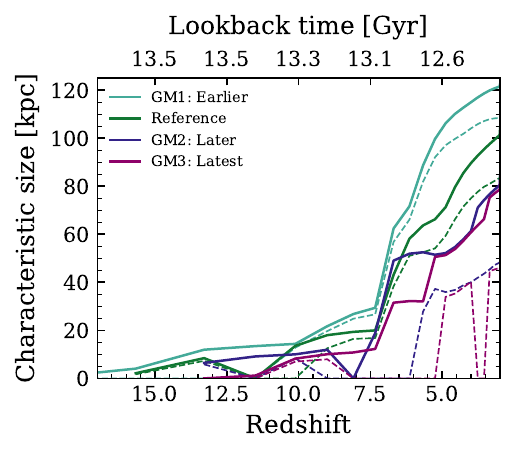}
    \caption{Characteristic size of volumes where the gas [Fe/H] exceeds $-4$ (filled lines) and $-2$ (dashed line) around the main progenitor. Simulations where the galaxy growth is concentrated (e.g., \texttt{GM1: Earlier}) have volumes that systematically grows faster compared to those where star formation is split between many halos.}
    \label{fig:enrichment_size}
\end{figure}

Because our systems have similar present-day halo masses, differences in assembly history lead to markedly different modes of star formation and chemical enrichment. Systems that assemble early experience rapid growth in stellar mass concentrated in a dominant central halo. In contrast, late assembly results in star formation across multiple smaller halos that subsequently merge. At the two extremes of our suite we identify two distinct regimes, primarily distinguished by the number of progenitors that contribute significant fractions of the final stellar mass. We emphasize that these regimes should not be interpreted as a universal classification of dwarf galaxy formation. Rather, they represent two limiting cases within our controlled set of simulations that provide a useful framework for interpreting how differences in assembly history may manifest in observable stellar populations. Our examples of these regimes, \texttt{GM1: Earlier} as a single-progenitor system and \texttt{GM3: Latest} as a multi-progenitor system, illustrate how assembly histories dominated by one or several major contributors can lead to qualitatively different chemical signatures. Because star formation is the only source of heavy elements in our simulations, these different regimes lead to distinct spatial patterns of chemical enrichment in the surrounding medium.

Figure~\ref{fig:halo_projection_map} illustrates the different regimes of star formation and iron enrichment, centered around the main progenitor galaxy in each simulation at three representative redshifts. In \texttt{GM1: Earlier}, star formation is strongly centralized, as shown in the left panel by marking locations of recent star formation in red. \texttt{GM3: Latest} (right panel) contrasts with this, instead showing recent star formation distributed across many different halos. As a result, single-progenitor systems show metals being injected primarily into the surroundings of the central halo, especially at earlier times. For multi-progenitor systems, metal production occurs across numerous halos. These patterns are shown using white contours placed over the projected gas density in Fig.~\ref{fig:halo_projection_map}. 

These differences are also reflected in the amount of iron and its spatial extent around the main progenitor galaxy. In Fig.~\ref{fig:enrichment_size} we show the evolution of the characteristic radius containing enriched gas. We estimate this radius as $r_{\rm char} = (3V/4\pi)^{1/3}$, where $V$ is the total volume within the isocontour surface around the main progenitor where [Fe/H] exceeds -4 and -2. Single-progenitor systems show rapid expansion of this radius, reflecting sustained star formation and cumulative outflows from a central galaxy. In the multi-progenitor regime, enrichment proceeds more gradually as heavy elements are produced in several lower-mass halos whose outflows are weaker and spatially fragmented. As a result, the characteristic radius grows slowly at early times with a mode of rapid increases only after multiple enrichment sites begin to overlap. 

The efficiency with which metals are retained or expelled is different between the two regimes. In early-forming systems, concentrated star formation rapidly produces enriched material, but deeper potential wells hinder its escape, in particular for $M_{200}>10^7\,{\rm M}_{\odot}$ \citep{Mead+2025a}. This allows material to be more efficiently retained or recycled through galactic fountain flows. In contrast, fewer metals are produced in low-mass halos and these metals are more easily lost to the surrounding medium due to shallow potential wells. These effects hinder internal chemical enrichment in multi-progenitor systems in comparison to single-progenitor systems. Furthermore, it is not clear how easily pre-enriched material reaches the centers of nearby galaxies, where star formation is taking place. Fully exploring the in- and outflows, retention, and metal loading factors in all halos in our simulation is beyond the scope of this work, so we leave that to future work.

Finally, hierarchical assembly introduces an additional channel for chemical diversity through late accretion of pre-enriched satellites. In our multi-progenitor systems, stars and gas from separate halos are incorporated into the main galaxy at later times, potentially bringing chemically distinct populations. However, external enrichment from the surrounding environments may partially erase these signatures in gas. If progenitors enrich the environment as a collective, chemically distinct populations in satellite systems might be blurred by this effect. Therefore, determining whether these processes preserve or erase distinct chemical signatures is key to interpreting the elemental fossil record of galaxy assembly.

\section{Observable signatures}\label{sec:stellar_properties}

We now examine signatures in the stellar populations that are observable at present day and ask whether these can distinguish different assembly histories. In particular, we focus on stellar ages, chemical abundances, and stellar kinematics. In previous work, we investigated how the derived galaxy properties depends on the number of observable members by selecting stars brighter than magnitude $23$ in the $r$-band at a distance of $200\,{\rm kpc}$ \citep[][also in Fig.~\ref{fig:feh_MdynLV}]{Andersson+2025}. While such a criterion is appropriate for photometrically derived quantities, the feasibility of spectroscopic measurements, such as detailed chemical abundances and stellar ages, depends more strongly on the specific observational setup and target selection. Rather than adopting a fixed magnitude limit, we therefore perform our analysis using the $100$ brightest stars in each galaxy (ranked by $r$-band magnitude). As discussed in Appendix~\ref{appendix:mdf}, this sample size is sufficient to recover the underlying distributions while remaining representative of the best-studied ultra-faint dwarf galaxies \citep[e.g.,][]{Luna+2025}.

In addition, when deriving the mean [Fe/H] (shown in Fig.~\ref{fig:feh_MdynLV}) and stellar brightness from isochrones,\footnote{We used the isochrones from \textsc{parsec} \citep{Bressan+2012}, see Appendix A in \citet{Andersson+2025} for further details.} we exclude all stars with $\mathrm{[Fe/H]}<-4$. These stars may carry signatures of Population~III enrichment \citep{Jeon+2017}, a process that is not included in our simulation. This reduces the number of stars in our analysis, particularly in the latest-forming galaxies where a significant portion of the stars are extremely metal-poor. We address how these stars impact the mass-metallicity relation in Chaturvedi et al. (subm.) and defer an investigation of the impact of Population~III enrichment on the observability of these systems to future work. Finally, we distinguish between stars formed in situ (i.e., within $R_{200}$ of the main progenitor halo) and those accreted ex situ from merging satellites.

\subsection{Chemical evolution of the stellar population}\label{subsec:chemical_evolution}

\begin{figure*}
    \centering
    \includegraphics[width=\linewidth]{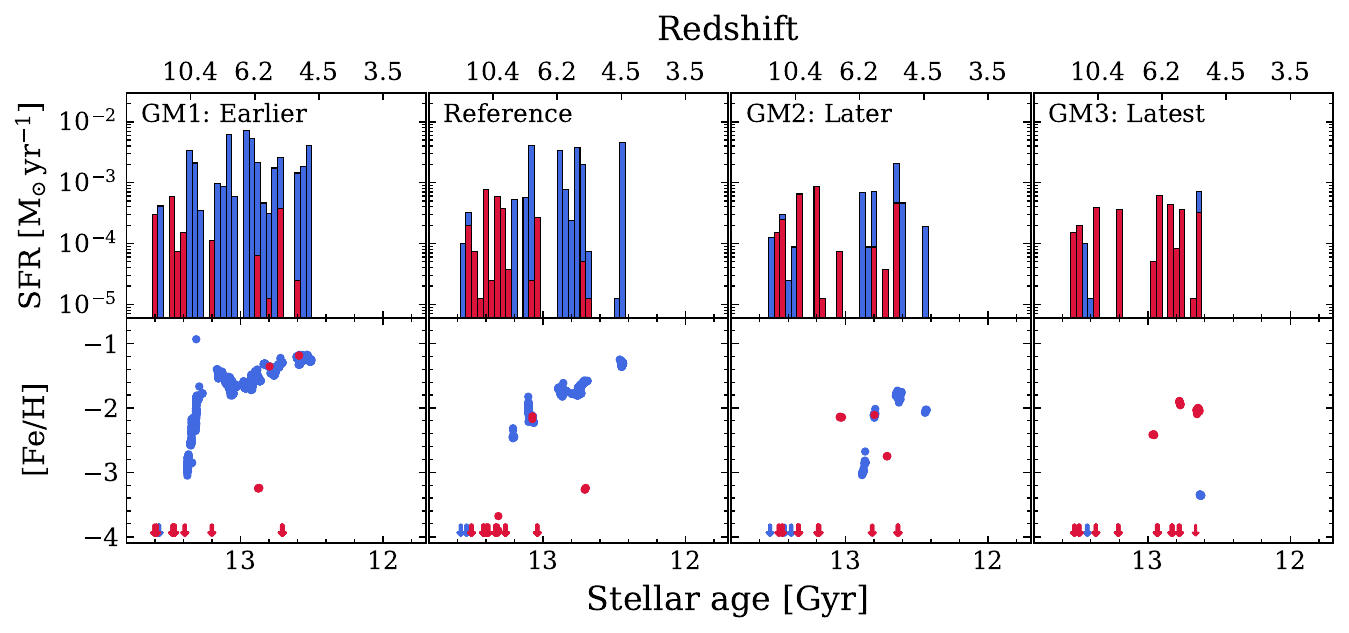}
    \caption{Star formation rate (top) and stellar [Fe/H] (bottom) as a function of stellar age at redshift zero (formation redshift is shown on the top axis). Each column corresponds to a different simulation. In the top panels, the blue curve shows the total star formation rate, while the red curve shows the contribution from accreted stars, plotted on top of the total. The in situ star formation rate is therefore given by the difference between the blue and red curves. In the bottom panels, all stars are shown in blue, with accreted stars overplotted in red. Consequently, blue points not covered by red correspond to in situ stars. Stellar [Fe/H] values have not been convolved with the 0.05 dex uncertainties applied in other figures. Stars outside the plotted abundance range are indicated by arrows pointing toward their values. Stars that are outside the range in the bottom panel are indicated by arrows pointing toward their value. We consider these abundance ratios possibly spurious due to missing Pop.~III physics (see text for details).}
    \label{fig:iron_sfr}
\end{figure*}

We begin by examining the chemical enrichment record of the stellar populations, focusing on differences between the single- and multi-progenitor regimes. Figure~\ref{fig:iron_sfr} shows the star formation rate (top row) and [Fe/H] (bottom row) as a function of stellar age at redshift zero for each simulation. Note that for Fig.~\ref{fig:iron_sfr}, we did not remove any stars based on magnitude, instead showing all stars from each simulation. Furthermore, stars below the [Fe/H] limit at $-4$ are indicated as arrows placed at the bottom in the lower panel indicating their formation time. Together, these quantities trace how star formation histories are linked to chemical evolution, which ultimately shapes the stellar populations observable at the present day.

At fixed present-day halo mass, early forming systems produce significantly larger populations of stars compared to late-forming ones, a difference primarily driven by in situ star formation.
In fact, late-forming systems form a large fraction of their stars in external progenitors that are later accreted. While the majority of star formation in these systems is associated with accreted material, the total mass of accreted stars is only a factor of two greater in \texttt{GM3: Latest} than in \texttt{GM1: Earlier}. The resulting in situ fractions are $0.96$, $0.89$, $0.60$, and $0.15$ for \texttt{GM1: Earlier}, \texttt{Reference}, \texttt{GM2: Later}, and \texttt{GM3: Latest}, respectively.

As a result of high in situ-star formation rates, stellar populations in the single-progenitor regime (represented by \texttt{GM1: Earlier}) exhibit a relatively continuous enrichment history. In this system, [Fe/H] increases monotonically as successive generations of stars form from progressively enriched gas. Around [Fe/H]$=-2$ in \texttt{GM1: Earlier}, we note a break in the enrichment history, after which the rise in [Fe/H] becomes more gradual. The \texttt{Reference} simulation only exhibits this later, slower phase of enrichment. Notably, \texttt{GM1: Earlier} has a halo mass that exceeds $10^{7}\,{\rm M_{\odot}}$ during the initial sharp rise in [Fe/H], a mass above which galaxies have been shown to start retaining metals \citep[][]{Mead+2025a}.

The multi-progenitor regime shows markedly different patterns. In these systems, star formation occurs in several smaller halos that later merge to form the final galaxy. The resulting enrichment history is therefore fragmented, consisting of multiple stellar populations with distinct [Fe/H] signatures that reflect their independent chemical evolution histories. Accreted stars are particularly prominent in this case and can even reach higher metallicities than stars formed in situ within the main progenitor. Consequently, stellar ages and metallicities in multi-progenitor systems do not trace a single coherent enrichment track but instead reflect the superposition of many independent star formation events. In the most extreme case, \texttt{GM3: Latest}, it becomes difficult to identify any single coherent chemical track at all.

\subsection{Metallicity distribution functions}

\begin{figure*}
    \centering
    \includegraphics[width=0.975\linewidth]{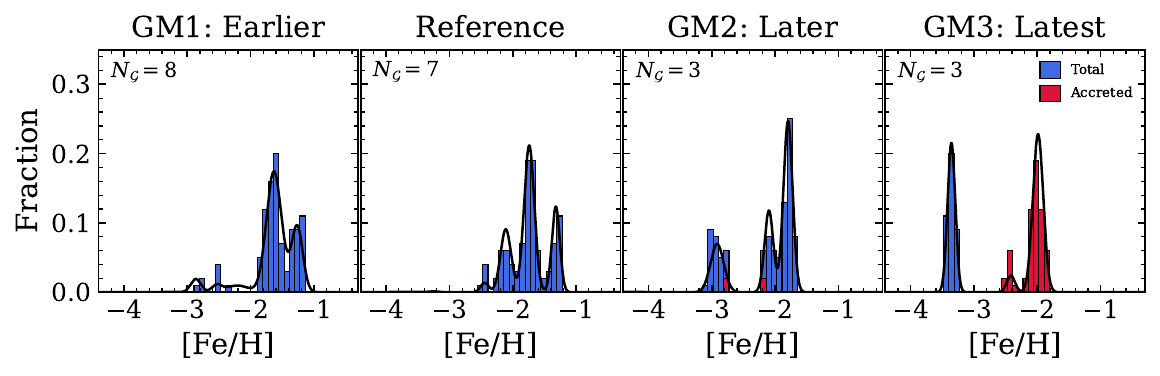}
    \caption{Metallicity distribution functions of each simulations as labeled, shown using [Fe/H] as a proxy for metallicity. Blue shows the distribution of the entire population and red the accreted stars. Histograms were calculated using only the $100$ brightest stars (see main text for details). The black line in each panel shows a GMM fitted to the entire data set with the number of Gaussians indicated by $N_\mathcal{G}$ in the upper left of each panel.}
    \label{fig:feh_mdf}
\end{figure*}

We now examine how the metallicity evolution of the different assembly histories is reflected in the present-day [Fe/H] distributions of the stellar populations of the galaxies. Figure~\ref{fig:feh_mdf} shows the metallicity distribution functions (MDFs) for each simulation, constructed from the $100$ brightest present-day stars, each convolved with a $0.05\,{\rm dex}$ Gaussian spread to avoid numerical issues associated with delta functions in the the Gaussian mixture model (GMM) applied later, and shown as histograms. This small spread is introduced as numerical smoothing and is not intended to represent a measurement uncertainty. We note that $0.05\,{\rm dex}$ is substantially smaller than typical observational uncertainties. As in previous figures, we color the histogram bins such that each bin shows both the total (blue) and accreted (red) populations. Because the MDF integrates the entire chemical evolution history of the galaxy, it provides a robust diagnostic that can be derived from a few tens to hundreds of stars in observed galaxies \citep[see, e.g.,][]{Fu+2023, Luna+2025}.

In the single-progenitor regime, the rapid and continuous enrichment of the ISM, as well as the ability of more massive systems to retain enriched material, leads to MDFs dominated by metal-rich stars. While several peaks can be discerned in the MDF, there is significant overlap between them. Applying a GMM (see Appendix~\ref{appendix:gmm}) reveals that the stellar population can be decomposed into multiple components. While most in situ star formation events result in a wide range of metallicity, some components reflect distinct star formation episodes which can masquerade as accretion events. In future work, we will explore methods to distinguish these peaks from accretion events and link them to specific burst. We note that the accreted populations in the single-progenitor regime are typically overshadowed by the number of stars formed in situ.

The multi-progenitor regime produces qualitatively different MDFs. Because stars form in several progenitor halos that evolve chemically at different rates, the final stellar population is a superposition of multiple chemical enrichment tracks. This leads to MDFs with distinct peaks, in most cases with little or no overlap. These peaks correspond to stars formed in different progenitor systems. However, the MDF alone does not uniquely identify accreted populations, as stars formed in situ and ex situ can occupy similar metallicity ranges. Furthermore, because of the low halo masses during the epoch when star formation is strongest, most halos are unable to retain enriched gas, and many stellar populations fall outside the [Fe/H] range that we consider.

Finally, we examine how observational limitations affect the ability to recover these features. In Appendix~\ref{appendix:mdf} we explore this further, showing that the overall MDF can be recovered provided that observed systems include at least $\sim50$–$100$ member stars. Below this threshold, the underlying [Fe/H] distribution becomes challenging to recover, particularly features that separate single- and multi-progenitor systems.

While MDFs provide a summary of the chemical enrichment history, they do not fully capture the complexity that assembly history imposes on the final stellar populations. In particular, stars formed in different progenitor systems may occupy similar metallicity ranges, making it difficult to identify accreted populations. In the following section we examine abundance planes, focusing on [$\alpha$/Fe] as a function of [Fe/H], which help reveal the chemical diversity of dwarf galaxies assembled from multiple star formation episodes in several progenitor halos.

\subsection{Abundance trends as tracers of distinct stellar populations}

\begin{figure*}
    \centering
    \includegraphics[width=0.975\linewidth]{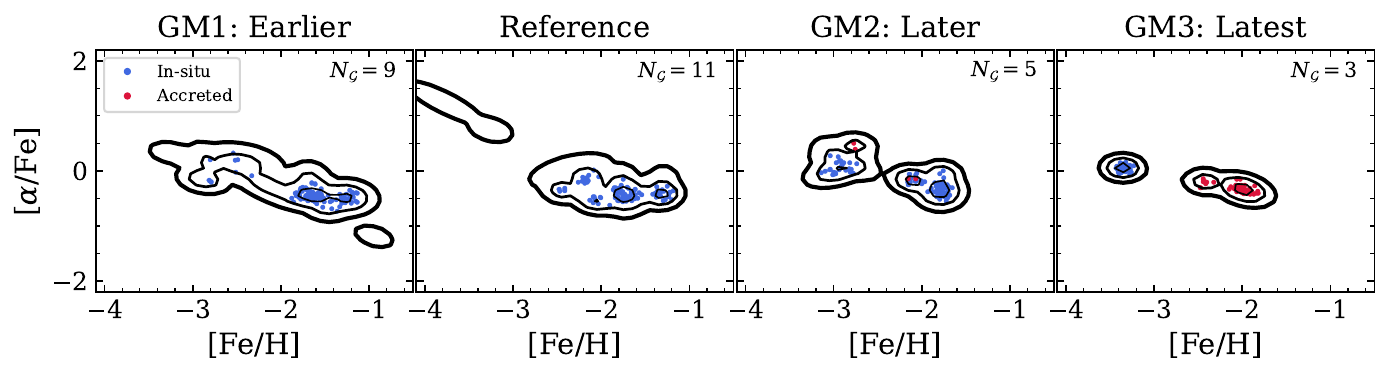}
    \caption{Abundance of $\alpha$ elements relative to iron as a function of iron abundance, showing observable in situ stars in blue and those accreted in red. Similarly to Fig.~\ref{fig:feh_mdf}, the total populations is shown by contours of a fitted GMM ($1$, $2$, and $3\,{\sigma}$ in increasing line width, and $N_{\mathcal{G}}$ indicating the number of Gaussians).}
    \label{fig:afefeh}
\end{figure*}

To disentangle populations that overlap in the MDFs, we investigate signatures in the $\alpha$-to-iron abundance ratios as an additional chemical dimension. We define [$\alpha$/Fe] by combining the available elements associated with $\alpha$-process nucleosynthesis,
\begin{equation}
\left[\alpha/\rm Fe\right] = \frac{1}{4}\left(\left[{\rm C}/{\rm Fe}\right] + \left[{\rm O}/{\rm Fe}\right] + \left[{\rm Mg}/{\rm Fe}\right] + \left[{\rm Si}/{\rm Fe}\right]\right).
\end{equation}
In Fig.~\ref{fig:afefeh}, we show [$\alpha$/Fe] as a function of [Fe/H] for each simulation. As in the previous section, we use the same observable stellar populations, separate in situ and accreted stars, apply the $0.05\,{\rm dex}$ numerical smoothing to both [Fe/H] and [$\alpha$/Fe], and include fitted GMMs to quantify distinct populations.

Adding a second chemical dimension reveals structure that is not readily apparent in the MDF alone. In the single-progenitor regime, the majority of stars follow a relatively coherent enrichment sequence in the [$\alpha$/Fe]–[Fe/H] plane, reflecting the more continuous chemical evolution of a dominant progenitor halo. Nonetheless, some features at fixed [Fe/H] show clear separation in [$\alpha$/Fe], and the GMM decomposition indicates that many distinct stellar populations are present. These components reflect the bursty nature of star formation but typically lie close to the main trend. Note that both \texttt{GM1: Earlier} and \texttt{Reference} have stars with $[\alpha/{\rm Fe}]$ and [Fe/H] values which appear in the GMM at the $3\,{\sigma}$ level, however these stars are too faint to appear in our sample of the $100$ brightest stars. In fact, $\mathcal{O}(1000)$ stars must be observed before all features in the GMM are populated by points.

In contrast, multi-progenitor systems produce abundance patterns with distinct features and little overlap, similar to the MDFs. This is expected, since star formation occurs in many different halos. However, the number of distinct features remains similar to those found in the MDFs. As in the single-progenitor regime, accreted material is clearly separated for all populations except one. However, in the multi-progenitor regime these populations are separated at the $1$–$2\,\sigma$ level in the GMM, which means that accretion signatures should be detectable with significantly smaller samples than in single-progenitor systems.

The larger number of components detected in the single-progenitor regime compared to the multi-progenitor regime is a result of the bursty nature of star formation. Furthermore, because late-forming systems correspond to our multi-progenitor representatives, these systems suffer from lower metal retention in low-mass satellite halos. This results in many stellar populations at extremely low metallicities and somewhat reduces the effective chemical dynamic range available to separate populations.

Finally, we note that the qualitative behavior described above is not unique to $\alpha$ elements. In fact, combining elements in this fashion somewhat reduces discriminating power and limits the number of detectable stellar populations. In Appendix~\ref{appendix:xfefeh}, we therefore show the abundance planes for all elements tracked in our simulations. While the detailed distributions vary depending on the nucleosynthetic origin of each element, and is more sensitive to the uncertainties in the underlying model \citep[e.g, yields, stellar rotation, and SNIa delay-time distributions][]{Andersson+2026}, the same broad trends remain: single-progenitor systems show a dominant enrichment sequence, while multi-progenitor systems exhibit clearly separated populations. We also find that many features that overlap in the [$\alpha$/Fe]–[Fe/H] plane become distinct and are easily detected by the GMM.

\subsection{Spatial distribution of in situ and ex situ stars}

\begin{figure*}
    \centering
    \includegraphics[width=0.975\linewidth]{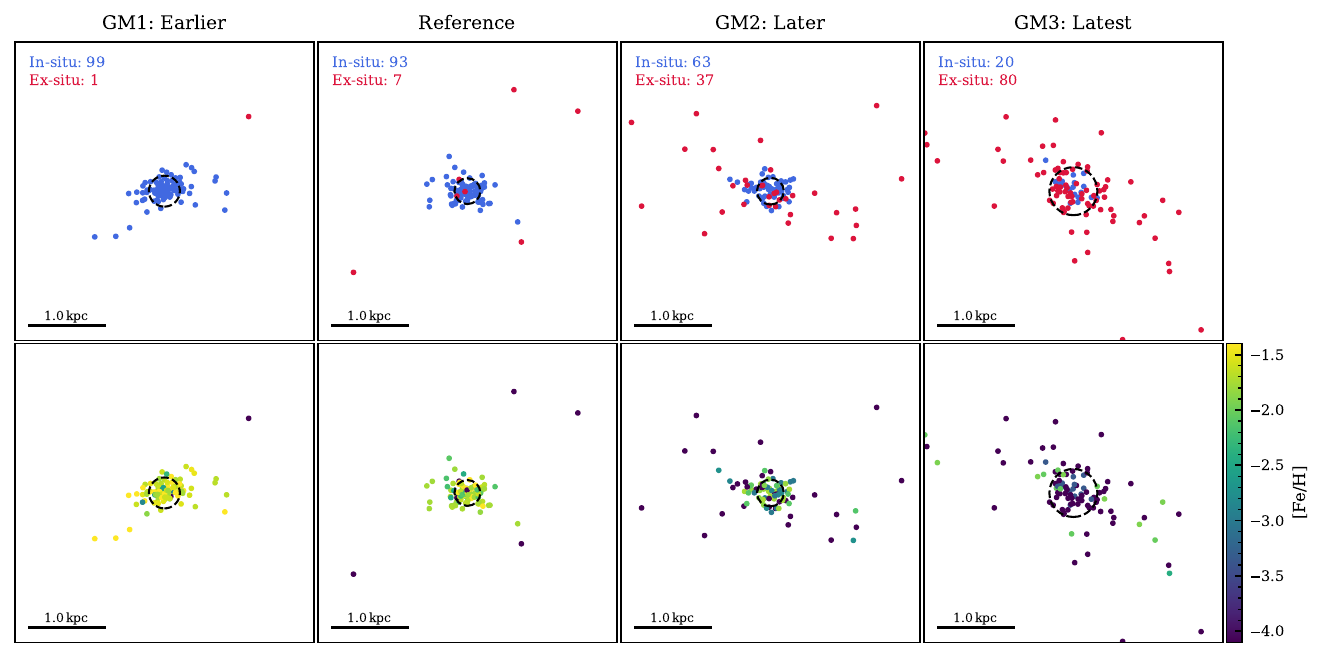}
    \caption{Spatial distribution of observable stars in the final snapshot ($z=0$), with colors indicating in situ (blue) and accreted (red) populations in the top row, while the bottom row shows metallicity as indicated by the color bar. {Out of the $100$ brightest stars, the numbers of in situ and ex situ stars are indicated in the top left corner of the upper panels. Scale is shown in the bottom left of each panel, and the half-light radius is indicated using black dashed line}.}
    \label{fig:star_positions}
\end{figure*}

Finally, we examine the spatial distributions of the stellar populations to determine how assembly history is reflected in the present-day structure of the galaxies. In Fig.~\ref{fig:star_positions} we show projected positions of observable stars in each simulation. The top row highlights the origin of the stars, separating stars into in situ (blue) and accreted (red). In combination, these maps show how hierarchical assembly shapes the spatial and chemical structure of the final galaxy.

Accreted stars are typically distributed over large radii compared to the in situ population, forming an extended stellar component surrounding the central galaxy. This extended halo drives differences in the half-light radius for early versus late forming galaxies \citep{Rey+2019, Andersson+2025}. Furthermore, the spatial distribution of accreted stars is broadly similar outside $100\,{\rm pc}$ across all simulations. If accreted stars have retained their dynamical imprint until present day, comparable spatial distributions would imply similar orbital energies of mergers in the different simulations. A first inspection of our data suggests that this is not the case, and instead we speculate that the accreted populations have experienced significant dynamical evolution. Fully exploring this is beyond our scope and we leave this to future work. We also note that our galaxies are selected to be isolated, and tides are not imposed by any nearby large galaxy by design.

In contrast, stars formed in the main progenitor are strongly concentrated to the center of the galaxy. Because the central region is the primary site of star formation, in situ stars are also able to retain their dynamical imprint over more than $10\,{\rm Gyr}$ of dynamical evolution. Qualitative this behavior is common to all simulations; however, in the multi-progenitor regime, the half-light radius estimate is significantly affected by the spatially extended accreted population that makes up a larger fraction of the total stellar population.

The metallicity distributions further illustrate how the spatial distribution of different populations is related to the chemical evolution of the galaxy. In most simulations, metal-poor stars are preferentially found at large radii and are often associated with accreted material, whereas the central regions are typically metal-rich. Single-progenitor systems show hints of metallicity gradients that we will explore in future work (Riley et al., in prep.). In the most extreme multi-progenitor system (\texttt{GM3: Latest}) though, the most metal-rich stars are found on wide orbits because they belong to the accreted population. While such systems might be uncommon, late incorporation of chemically evolved systems should be accounted for when interpreting observed data.

In summary, while chemical abundance patterns encode the independent enrichment histories of different progenitors, the spatial distribution broadly preserves the overall dynamical imprint of assembly, i.e., concentrated in situ populations and more extended accreted populations. Taken together, these properties provide a promising way to distinguish single- and multi-progenitor systems. However, our simulations show signs that the dynamical imprints have evolved to some extent. Furthermore, large stellar samples are required, and reliably identifying member stars remains challenging, particularly for accreted stars that can reside far outside the galaxy’s half-light radius.

\subsection{Implications for Local Group dwarf galaxies}\label{subsec:observations}

Our comparison between the single- and multi-progenitor regimes is based on galaxies with comparable dynamical masses at the present day, but with growth histories that result in different stellar masses. In our case, this is achieved by design using genetically engineered initial conditions \citep{Roth+2016, Rey&Pontzen2018, Stopyra+2021}. This difference is quantified by mass-to-light ratio, as shown in Fig.~\ref{fig:feh_MdynLV}. Nearby dwarf galaxies with comparable dynamical mass but spanning a range of total magnitudes include Ursa Minor, Ursa Major I, Eridanus II, Draco, Leo II, and possibly Bootes I, although it has a relatively low dynamical mass. With the exception of Leo II, these galaxies show predominantly old stellar populations, consistent with either quenching during reionization \citep[$12\,{\rm Gyr}$ ago][]{Okamoto+2008, Shetrone+2009, Read&Erkal2019, Gallart+2021, Simon+2021} or other environmental effects ($10\,{\rm Gyr}$ ago) in the case of Ursa Minor \citep{Sato+2025} and Draco \citep{Aparicio+2001}.

Nearby dwarf galaxies exhibit a range of properties that qualitatively resemble both the single- and multi-progenitor regimes identified in our simulations. While Ursa Major I could be a good candidate for the multi-progenitor regime, very limited spectroscopic data currently exist \citep{Garofalo+2013, Geha+2026}. More extensive observations are available for Eridanus II, a highly dark-matter-dominated system with a mass-to-light ratio of $\approx100$ \citep{Pace2025}, reminiscent of our \texttt{GM2:Later} simulation. In contrast to single-progenitor systems, Eridanus II is expected to retain clearer signatures of hierarchical assembly in its stellar abundances. Photometric metallicities reveal a broad distribution extending to ${\rm [Fe/H]}\approx-4$ \citep{Zoutendijk+2020, MartinezVazquez+2021, Fu+2022}, while spectroscopic measurements suggest multiple peaks of comparable height in the MDF \citep{Li+2017}, albeit based on only 16 stars. Its high ellipticity and the presence of a central star cluster \citep{Simon+2021} further provide independent evidence consistent with a past accretion event \citep{Orkney+2022, Goater+2024, Andersson+2025}.

Ursa Minor also appears to be a promising candidate for the multi-progenitor regime. Multiple chemo-dynamically distinct stellar populations have been identified in this galaxy \citep{Pace+2020}, a result that is difficult to reconcile with simple one-zone chemical evolution models \citep{Cohen+2010}. Furthermore, Ursa Minor has a high central velocity dispersion, and fully understanding its assembly history is important for models of dark matter \citep{Read+2019, Muni+2025, Pascale+2026}. Although Bootes I has a comparatively low dynamical mass, the relatively large spectroscopic samples now available \citep{Sandford+2026} make it one of the best-studied systems in this mass range, and its chemical abundance patterns have likewise been interpreted as evidence for an accretion event \citep{Frebel+2016}. Further into the ultra-faint regime, \citet{Luna+2025} recently reported metallicities for 167 stars in Reticulum II, representing the largest spectroscopic sample obtained for an ultra-faint dwarf galaxy to date. They identified a possibly multi-modal MDF and argued that the distinct populations likely originated from two separate episodes of star formation rather than mergers between multiple halos, based on the spatial mixing of the stellar populations. Furthermore, the metal rich mode appears to be r-process enriched \citep{Luna+2025}. Regardless of the underlying origin, these observations demonstrate that chemically distinct populations are present in at least some ultra-faint dwarfs.

In contrast, Draco exhibits characteristics that are more reminiscent of the single-progenitor regime. It has a mass-to-light ratio comparable to \texttt{Reference} (see Fig.~\ref{fig:feh_MdynLV}) and a predominantly old stellar population (\citealt{Aparicio+2001}, but see also \citealt{Orban+2008}). Extensive spectroscopic studies have revealed a relatively smooth MDF and coherent abundance trends that can largely be reproduced by simple chemical evolution models \citep{Kirby+2011, Kirby+2013, Cohen&Huang2009}. Although substructure cannot be ruled out, Draco lacks the clear evidence for multiple chemically distinct populations seen in systems such as Eridanus II, Ursa Minor, and Reticulum II. In the context of our simulations, Draco therefore provides an observational example of a galaxy whose chemical properties are qualitatively consistent with enrichment dominated by a single progenitor.

While the dwarf galaxy Sculptor is roughly two magnitudes brighter than the brightest galaxy in our sample, it provides a useful benchmark for chemical evolution. With a relatively low mass-to-light ratio ($\approx15$; \citealt{Lokas2009}), Sculptor is one of the best-studied dwarf galaxies, with photometric metallicities measured for more than $10^3$ stars \citep{Barbosa+2025} and multi-element spectroscopic abundances available for $\sim100$ stars \citep{Hill+2019}. Its stellar population follows a well-defined chemical evolution sequence \citep{Hill+2019}, with an MDF consistent with gradual enrichment from [Fe/H]$\approx-3$ to a sharp cutoff near [Fe/H]$\approx-1$. Using a methodology similar to ours, \citet{Ting&Ji2025} showed that GMMs applied to both the MDF and the [Mg/Fe]–[Fe/H] plane can recover signatures of bursty star formation, demonstrating that such statistical techniques can extract information about galaxy evolution even in systems that do not exhibit the chemically distinct populations expected in our multi-progenitor regime.

\section{Summary and conclusions}\label{sec:conclusion}

We explored how galaxy assembly history shapes the chemical evolution of low-mass galaxies and creates diversity among these systems. We analyzed four simulations of galaxies engineered to have similar present-day halo masses ($\sim1.4\times10^9\,{\rm M}_{\odot}$) but systematically varying growth histories. In this way, we isolate the role of hierarchical assembly in determining the elemental record of the final galaxies. Despite the comparable final halo masses, our simulations exhibit markedly different evolution of chemical and structural properties, demonstrating assembly history as a key driver of diversity among low-mass galaxies.

In order to maintain similar present-day halo mass, systems that assemble early evolve primarily through stellar mass growth in a single dominant progenitor, while late-forming systems have star formation divided between multiple halos with comparable mass. These two modes of assembly lead to qualitatively different star formation and enrichment histories, which in turn produce distinct signatures in present-day stellar populations. With this in mind, we define two distinct formation regimes; single- and multi-progenitor, represented by the extremes of early- and late-forming galaxies, respectively.

In the single progenitor regime, star formation is concentrated in a central halo that grows rapidly at early times. The resulting in situ star formation produces a relatively continuous chemical enrichment history, characterized by a rapid increase in [Fe/H] of newly formed stars and higher final mean metallicity. In contrast, multi-progenitor systems form through star formation distributed across many smaller halos that later merge to assemble the final galaxy. In this regime, each progenitor evolves along its own enrichment track prior to merging, often showing enrichment from a single burst of star formation. The resulting elemental record therefore represent a superposition of of several stellar populations with distinct chemical histories.

In summary, we draw the following conclusions.
\begin{enumerate}
    \item With concentrated star formation and rapid enrichment, single progenitor systems create regions of enriched material that grow rapidly around the center of the galaxy, often enclosing nearby satellites. In multi-progenitor systems, smaller regions of enriched material surround each progenitor, resulting in patchy enrichment of the environment, often without significantly enriched material reaching surrounding halos.
    \item Single progenitor systems have small fractions of accreted stars and show single coherent chemical evolution tracks with some substructure related to their bursty star formation histories. In contrast, multi-progenitor systems have significantly larger numbers of accreted stars, resulting in elemental abundance features that are clearly separated and often contain comparable numbers of stars. 
    \item Metallicity distributions provide a powerful diagnostic that can provide hints of the underlying assembly history. In particular, the presence of multiple peaks with comparable height can indicate that a galaxy belongs to the multi-progenitor regime. For single-progenitor systems, the rapid enrichment instead results in stellar populations with the majority of the stars in a few broad, overlapping peaks at high metallicity, and only a small fraction of stars at the low metallicity end. Nonetheless, multi-element abundance measurements can significantly improve the ability to separate these stellar populations.
    \item Accreted stars are typically on wide orbits, while in situ stars are centrally concentrated, even after $10\,{\rm Gyr}$ of dynamical evolution post star formation. In agreement with previous work \citep{Rey+2019, Andersson+2025}, this results in systematically larger half-light radii for multi-progenitor systems where the accreted population makes up a larger fraction of the total population.
    \item Chemical signatures of assembly history are typically well-separated compared to uncertainties from chemical enrichment models \citep{Andersson+2026}. Statistical tools, such as GMMs, provide a useful way to quantify different stellar populations, and are often able to find features that are difficult to distinguish visually. 
    \item However, the ability to determine assembly history from chemical abundances depends strongly on the number of observed stars. Our analysis suggests that samples of at least $\sim50-100$ member stars are required to reliably recover the underlying chemical structure of the galaxy, and perhaps even larger samples in single progenitor systems where many features hold relatively low numbers of stars. 
\end{enumerate}

Our results highlight the need for caution when interpreting metallicity distributions and chemical abundance patterns in observed galaxies. In extreme multi-progenitor systems, accreted stars can produce prominent high-metallicity features containing relatively large numbers of stars. In such cases, metal-rich populations may not trace the chemical evolution of the central galaxy, but instead reflect the late accretion of chemically evolved satellites. This can complicate the interpretation of metallicity gradients and inferred enrichment histories. However, spatial information provides a key diagnostic: accreted populations are typically found on extended orbits. In our most extreme multi-progenitor case, the highest-metallicity stars are located at large radii, consistent with an accreted origin. 

\begin{acknowledgements}
    We thank the anonymous reviewer for constructive feedback that improved the quality of this work.
    
    EPA and M-MML acknowledge support from NASA ATP grant 80NSSC24K0935. M-MML also acknowledges partial support from NSF grant AST23-07950.

    APJ acknowledges support from NSF grant AST-2307599, NASA ADAP grant 80NSSC25K7570, and the Alfred P. Sloan Foundation.

    JIR acknowledges support from STFC grants ST/Y002857/1 and ST/Y002865/1.

    OA acknowledges support from the Knut and Alice Wallenberg Foundation, the Swedish Research Council (grant 2025-04892), the Swedish National Space Agency (SNSA grants 2023-00164 and 2025-00405), the LMK foundation, and eSSENCE, a Swedish strategic research programme in e-Science.

    AC acknowledges the support from UKRI Science and Technology Facilities Council (STFC) grant ST/Y509619/1.

    AP was supported by funding from the European Research Council (ERC) under the European Union's Horizon 2020 research and innovation programmes (grant agreement no.\ 818085 GMGalaxies)

    This work made extensive use of the dp191 and dp324 projects on the STFC-DiRAC ecosystem. This work was performed using the DiRAC Data Intensive service at Leicester, operated by the University of Leicester IT Services, which forms part of the STFC DiRAC HPC Facility (www.dirac.ac.uk). The equipment was funded by BEIS capital funding via STFC capital grants ST/K000373/1, ST/R002363/1, and STFC DiRAC Operations grant ST/R001014/1. This work used the DiRAC@Durham facility managed by the Institute for Computational Cosmology on behalf of the STFC DiRAC HPC Facility (www.dirac.ac.uk). The equipment was funded by BEIS capital funding via STFC capital grants ST/P002293/1, ST/R002371/1 and ST/S002502/1, Durham University and STFC operations grant ST/R000832/1.

\end{acknowledgements}

\bibliographystyle{aa}
\bibliography{ref}

\begin{appendix}

\section{Gaussian mixture model}\label{appendix:gmm}

Throughout this work, we used a Bayesian Gaussian Mixture (BGM) model to describe the underlying distribution of chemical elements in our simulations. This model assumes that the data is described by a number $N_{\mathcal{G}}$ of Gaussians $\mathcal{G}$, i.e., drawn from
\begin{equation*}
    f(x) = \sum_k w_k \mathcal{G}(x|\mu_k,\sigma^2_k),
\end{equation*}
where for component $k$, $w_k$ is the weight, $\mu_k$ is the mean, and $\sigma^2_k$ is the variance. Using the implementation from \textsc{scikit-learn} \citep{scikit-learn}, the BGM model fits the data to $N_{\mathcal{G}}$ normal distributions using a Dirichlet Process \citep{Blei&Jordan2006, Bishop2006}. Unlike traditional GMMs, BGM does not rely on an assumption of $N_{\mathcal{G}}$, instead inferring it from the data.

When performing the fitting process, a few parameters must be tuned to the specific data. We let each component have its own covariance matrix, adding a regularization of $0.001$ to remove noise from single points. We apply a weight prior of unity, and only consider components that explain at least $2\%$ of the data. We found that the model robustly predicted the distribution for $N_{\mathcal{G},{\rm max}}\geq20$, and needed up to $2000$ iterations to find a converged solution for a tolerance of $0.005$.

\section{Observable metal distribution functions}\label{appendix:mdf}

\begin{figure*}
    \centering
    \includegraphics[width=0.875\linewidth]{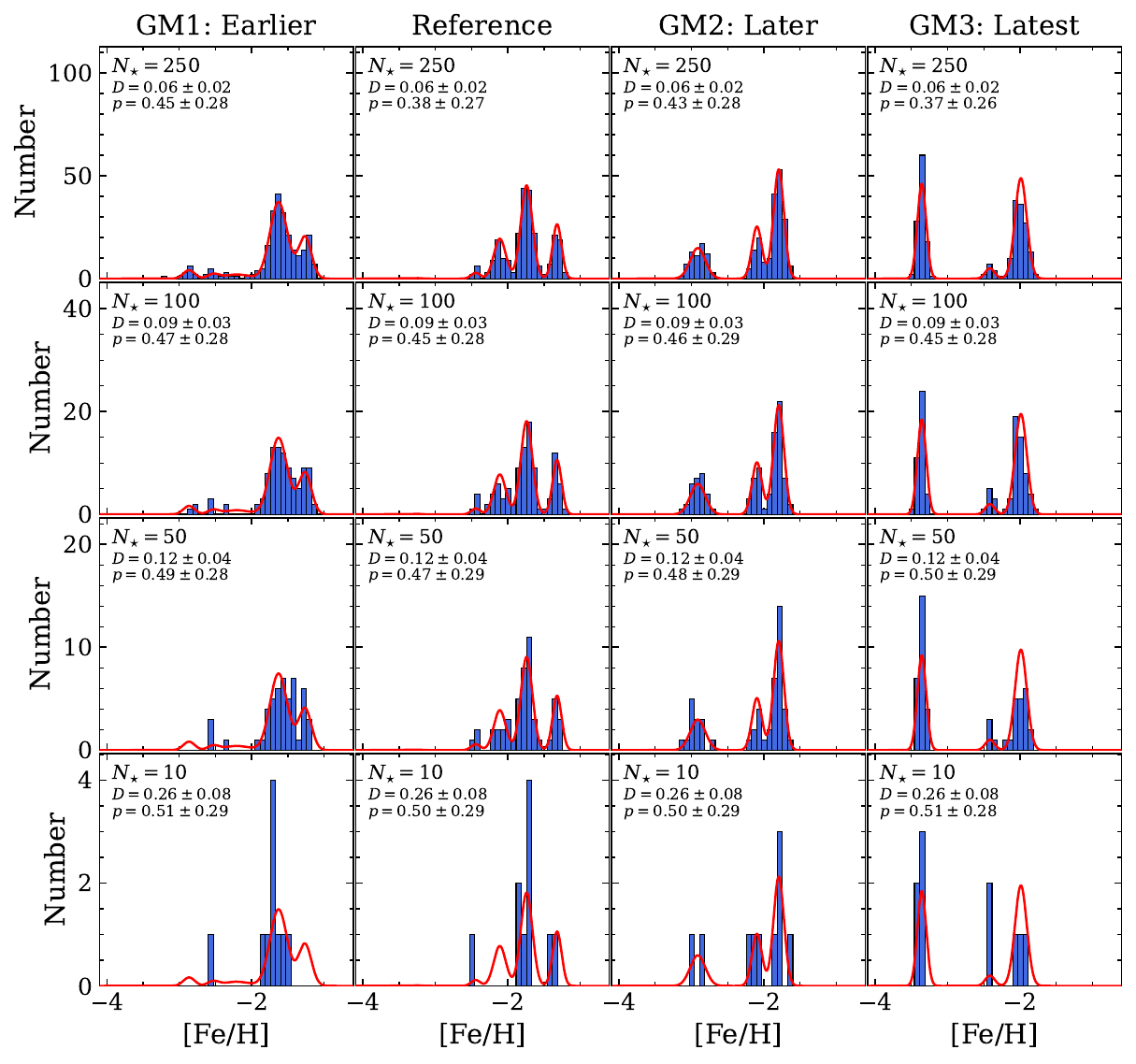}
    \caption{Metallicity distributions for each simulation by column, shown for the $N_{\star}$ brightest stars (indicated in the upper left corner of each panel). Each distribution is compared to the underlying distribution shown by GMMs in red lines. For each panel, we quantify whether the sample recovers the true distribution using a KS test ($D$ and $p$ values), see main text for details.}
    \label{fig:mdf_obs_limited}
\end{figure*}

We investigate whether different size samples of observed stars are able to reproduce the underlying distribution of [Fe/H]. To select the observed sample, we derive the magnitude of each star from mass, age, and metallicity using isochrones from \textsc{parsec} \citep{Bressan+2012}. A complete description of the derivation can be found in Appendix A of \citet{Andersson+2025}. We then select the brightest $N_{\star}$ stars based on $r$-band magnitude.

Figure~\ref{fig:mdf_obs_limited} shows representative distributions for $N_{\star}$ of $250$, $100$, $50$ and $10$. For each $N_{\star}$, we repeatedly performed a Kolmogorov–Smirnov (KS) test \citep{Massey1951} using $1000$ bootstrapped samples. To generate different samples, we recalculated the stellar magnitudes with small perturbations on stellar mass in very fine mass bins ($0.25\,{\rm M}_{\odot}$), while ensuring that the stars follow the same IMF. Note that stellar masses are randomly sampled in the simulation, and should be considered a source of statistical uncertainty. In addition, we add a uncertainty of $0.05\,{\rm dex}$ to our [Fe/H] measurements. We show the results of the KS statistic in the top left corner of each panel of Fig.~\ref{fig:mdf_obs_limited}, where $D$ is the largest difference between the observed and the underlying cumulative distributions, and $p$ tests the null hypothesis that the two samples are drawn from the same distribution. For each $D$ and $p$, the values show the mean and the error interval shows the standard deviation of the bootstrapped samples. Note that an increase in $D$ with decreasing $N_{\star}$ is consistent with the expected sampling fluctuations of the empirical cumulative distribution, expected to be on the order of $1/\sqrt{N_{\star}}$.

We find similar results between all four simulations. For $N_{\star}=250$, the empirical metallicity distribution of the brightest stars is very close to the fitted GMM, as indicated by $D\approx0.06\pm0.02$ and $p=0.3-0.4$. As $N_{\star}$ decreases towards $50$, the $D$ statistics show a moderate increase, while $p$-values remain large. At $N_{\star}=10$, the mock observed samples have distributions that are very coarse, therefore, despite $p$ being larger, the KS test has little constraining power. In summary, samples of roughly $50$–$100$ stars are required to reliably constrain the underlying metallicity distribution. 

\section{Individual abundances}\label{appendix:xfefeh}

\begin{figure*}
    \centering
    \includegraphics[width=0.975\linewidth]{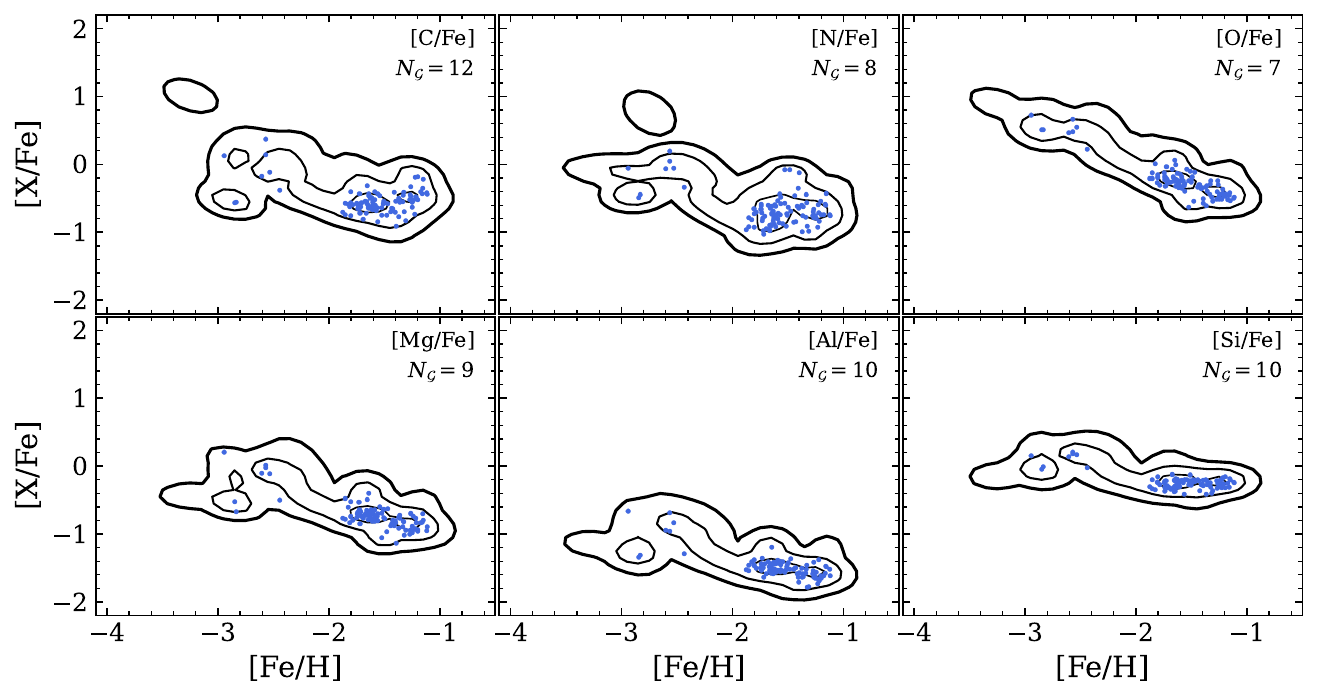}
    \caption{Relative abundance [X/Fe] (elements indicated in upper right corner) as a function of [Fe/H] for \texttt{GM1: Earlier}, showing observable in situ stars in blue and those accreted in red. The number of displayed stars are the same as in Fig.~\ref{fig:afefeh}, and the distribution of the total populations is shown in contours (increasing line width showing $1$, $2$, and $3\,{\sigma}$ of GMMs). The number of Gaussians is indicated by $N_{\mathcal{G}}$ in the upper right corner.}
    \label{fig:xfefeh_Mr12}
\end{figure*}

\begin{figure*}
    \centering
    \includegraphics[width=0.975\linewidth]{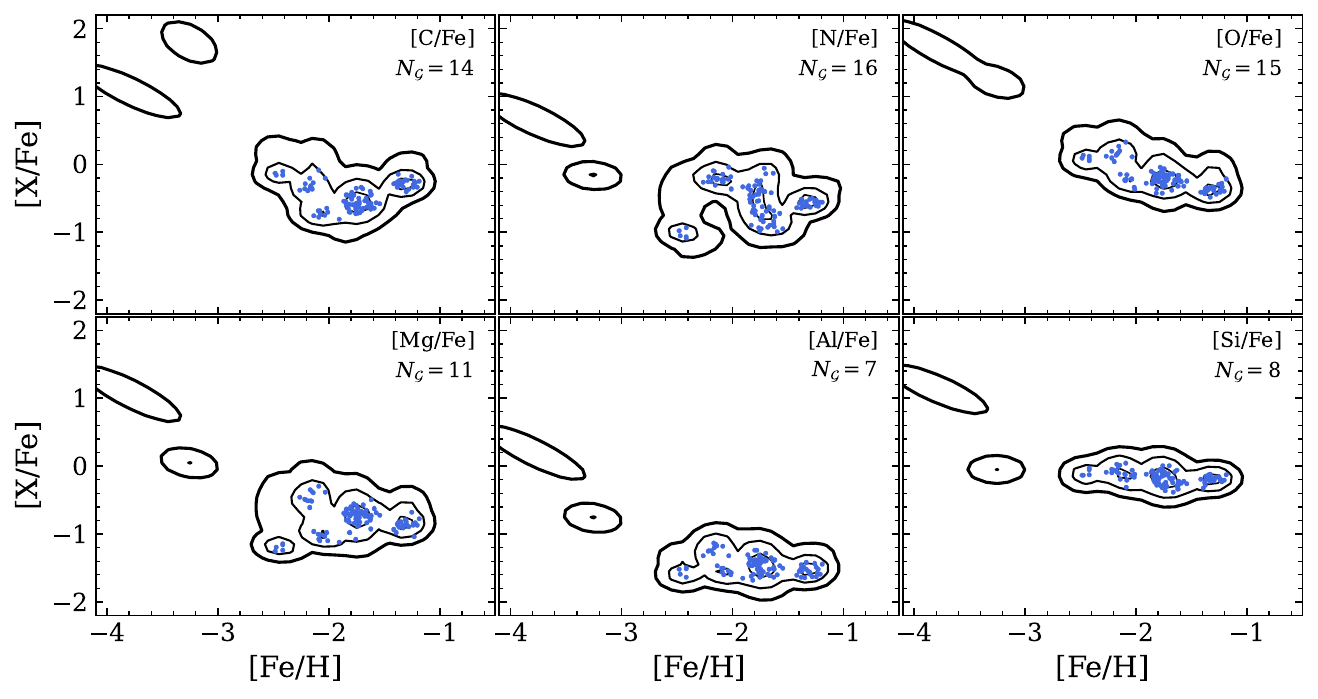}
    \caption{Same as Fig.~\ref{fig:xfefeh_Mr12} for \texttt{Reference}}
    \label{fig:xfefeh_Mr1}
\end{figure*}

\begin{figure*}
    \centering
    \includegraphics[width=0.975\linewidth]{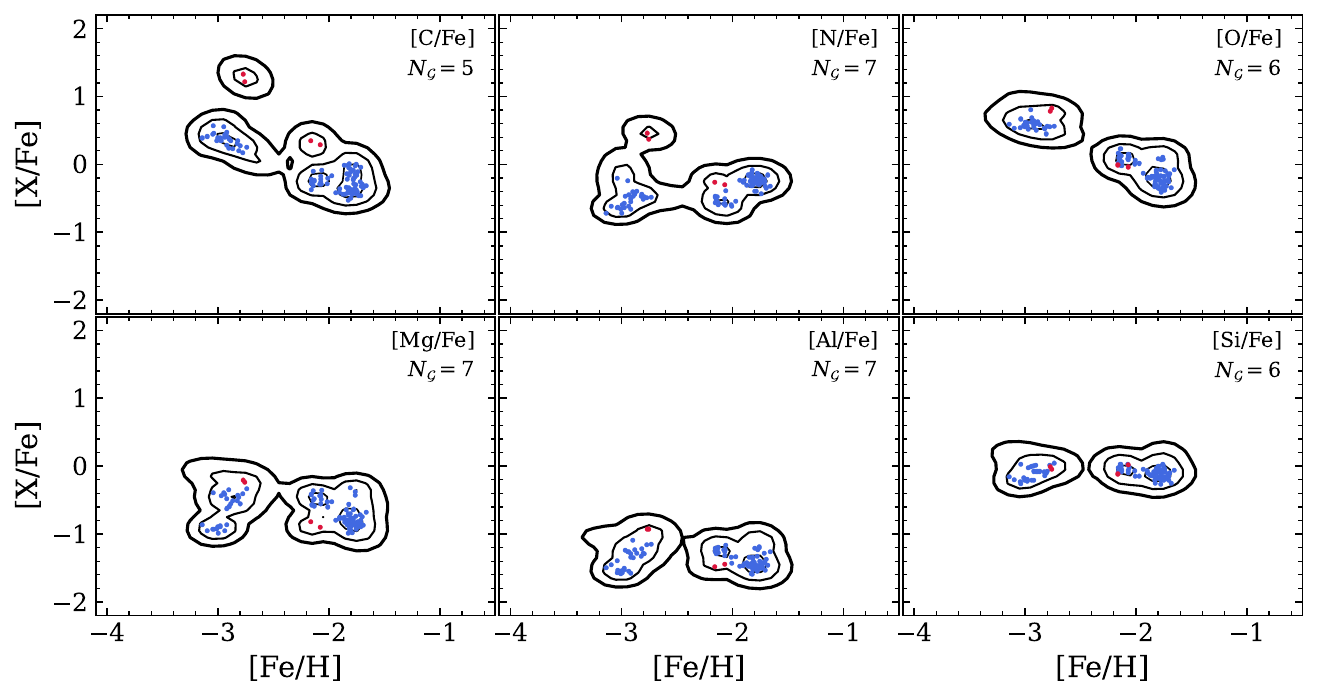}
    \caption{Same as Fig.~\ref{fig:xfefeh_Mr12} for \texttt{GM2: Later}}
    \label{fig:xfefeh_Mr03}
\end{figure*}

\begin{figure*}
    \centering
    \includegraphics[width=0.975\linewidth]{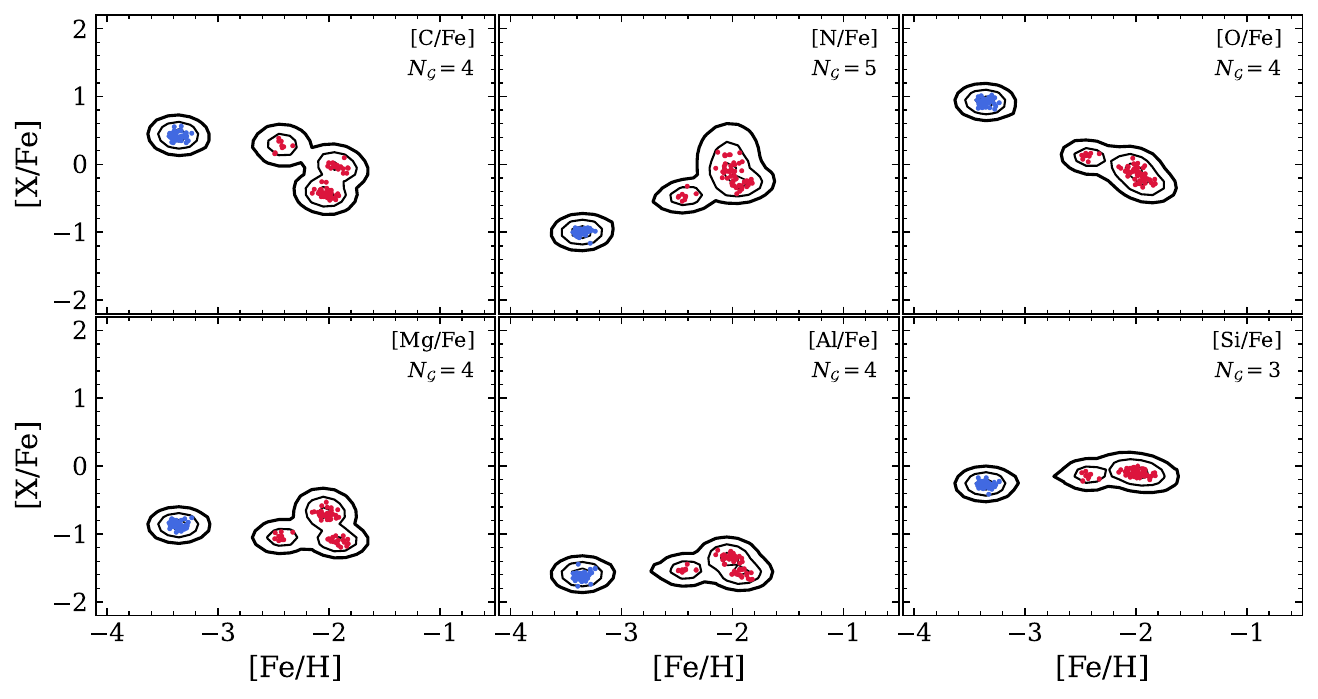}
    \caption{Same as Fig.~\ref{fig:xfefeh_Mr12} for \texttt{GM3: Latest}}
    \label{fig:xfefeh_Mr02}
\end{figure*}

In the main text, relative abundances of light elements are combined into [$\alpha$/Fe]. For completeness, we include the relative abundances of all traced elements in this section. Plotted similarly to [$\alpha$/Fe] in Fig.~\ref{fig:afefeh}, each simulation is shown in Fig.s~\ref{fig:xfefeh_Mr12}, \ref{fig:xfefeh_Mr1}, \ref{fig:xfefeh_Mr03}, and \ref{fig:xfefeh_Mr02} for \texttt{GM1: Earlier}, \texttt{Reference}, \texttt{GM2: Later}, and \texttt{GM3: Latest}, respectively. Individual elements show broadly consistent results compared to [$\alpha$/Fe], particularly for the light elements (C, N, O, Mg). To detect large numbers of structures, [N/Fe] is typically the best abundance ratio, although we note that a larger number of structures is detected for [C/Fe] in \texttt{GM1: Earlier}.

Furthermore, we find that many structures that overlap in some abundance planes (including [$\alpha$/Fe]) are separated in others. Notable examples of this behavior are the features linked to accreted material at high [Fe/H] in \texttt{GM3: Latest} and at [Fe/H]$\approx-2$ in \texttt{GM2: Later}. For the former, the accreted material show two structures in [Si/Fe], and three for the other elements, even showing little to no visual overlap between these features in [C/Fe] and [Mg/Fe]. For the latter, the feature clearly separates into two populations in [C/Fe], [N/Fe], [Mg/Fe] and marginally in [Al/Fe]; however, it appears as a single population in [O/Fe] and [Si/Fe].

\end{appendix}

\end{document}